\documentclass[aps,prl,twocolumn,groupedaddress,showpacs]{revtex4}
\usepackage{graphicx}
 
\begin{document}

 
\title{Quantum creep and quantum creep transitions in 1D
       sine-Gordan chains}


\author{Florian R. Krajewski}
\affiliation{
 Institut f\"ur Physik, WA 331, Johannes Gutenberg-Universit\"at Mainz, 55099 Mainz, Germany}

\author{Martin H. M\"user}
 \affiliation{Department of Applied Mathematics, University of Western Ontario,
              London, Ontario N6A 5B7, Canada}


\date{\today}

\begin{abstract}
Discrete sine-Gordon (SG) chains are studied with path-integral molecular 
dynamics. 
Chains commensurate with the substrate show the transition from collective
quantum creep to pinning at bead  masses slightly larger than
those predicted from the continuous SG model.
Within the creep regime, a field-driven transition from creep to
complete depinning is identified. 
The effects of disorder in the external potential on the chain's dynamics
depend on the potential's roughness exponent $H$, i.e.,
quantum and classical fluctuations affect the
current self-correlation functions differently for $H = 1/2$.
\end{abstract}


\maketitle


The discrete sine-Gordon (SG) chain, also known as 
Frenkel-Kontorova (FK) model~\cite{frenkel38,braun98},
is a generic model for the motion  of an elastic object composed 
of discrete degrees of freedom through an external potential.
To name a few  of its applications,
the SG model is used for the description of driven charge-density waves 
in solids, coupled Josephson junctions~\cite{floria96},
the sliding motion of an adsobed layer of atoms over a 
subtrate~\cite{muser03acp}, and most recently
electronic conductance in nanotubes~\cite{levitos03},
see also the reviews~\cite{ando98,zotos03} 
on electronic transport in one-dimensional structures.
A lot of attention has been devoted to the (discrete) classical
FK model both at zero and finite temperatures~\cite{braun98,floria96} and
the (continuum) quantum-mechanical 
SG model~\cite{zotos03,sklyanin79}.
However, less is known about the quantum-mechanical properties of the 
{\it discrete} quantum FK (QFK) model,
in particular about its {\it dynamical} properties.
Numerical approaches have lead to a clear
picture, how quantum fluctuations renormalize the thermal
equilibrium structure, but results are often limited to zero
(or small) external fields~\cite{hu99epl} or to variational 
approaches~\cite{ho01}. 
While quantum Monte Carlo (QMC) simulations 
yield numerically exact results for static properties, they only allow one to 
calculate small-frequency dynamical properties indirectly, i.e., conclusions
on the existence of a phonon gap are drawn by studying the temperature 
dependence of the internal energy~\cite{borgonovi89}.
In some cases, more dynamical information can be withdrawn from
QMC if the functional form of the low-energy spectrum is 
known~\cite{alvarez02}.

It is well established for the SG model that the effects due to 
thermal and those due to quantum fluctuations differ qualitatively.
Thermal fluctuations automatically lead to creep, while quantum fluctuations
do not.  At finite $T$,   kink 
anti-kink pairs will be activated and a small
external force will eventually be able to drive the pairs apart,
resulting in net mass transport~\cite{hanggi90}. 
If one assumes random interaction between the elastic object and 
the embedding system, thermal creep is also present
in higher dimensions, as shown in a sophisticated renormalization group 
(RG) study~\cite{chauve98epl}. 
Strictly speaking this implies that the pinning or the static friction force 
$F_s$ is zero and that
the kinetic friction force $F_k$ vanishes linearly with sliding 
velocity $v_0$ in the limit of $v_0 \to 0$, even thoug the proportionality 
coefficient $\gamma_{\rm eff} = \lim_{v_0 \to 0} F_k/v_0$
may be large.
In order for the 1D, {\it quantum} sine-Gordon model to
creep, it is not sufficient to have arbitrarily small quantum fluctuations,
but the effective masses $m$  (defined as density times period of
substrate potential)
must be less than a certain critical value $m_c$ (at fixed momentum cut-off
and fixed substrate strength)~\cite{sklyanin79}. 
For $m > m_c$, thermal fluctuations and/or finite external forces
are required to initiate mass transport. 
Recent RG studies~\cite{gorokhov02} suggest that $m_c$ is also finite
in higher dimensions if the elastic
manifold is pinned through an external random potential.

In many of the above mentioned cases, including the quantum SG model,
it is necessary to go beyond standard RG 
theories~\cite{chauve98epl,gorokhov02,kehrein99}, 
because otherwise the effect of creep or alternatively the effect of 
a vanishing excitation gap at zero wave vectors
might be artificially removed. 
It is thus desirable to have numerical
techniques which allow one to (i) verify the results of RG studies,
(ii) obtain  results beyond continuum approximations allowing an
accurate determination of critical values such as $m_c$ for discrete
systems, and (iii) obtain dynamical responses of the system
for arbitrary external forces. In this Letter,
we first intend to establish that (adiabatic) path integral 
molecular dynamics (aPIMD)~\cite{tuckerman93,cao96} 
is a well-suited technique to tackle the 
many-body 
dynamics of elastic manifolds moving through embedding systems. 
We will then study  the FK chain's
dynamics as a function of the external field. This analysis
includes
new results 
for the quantum dynamics of a FK chain with
various types of disordered substrates.

PIMD is a  method to calculate 
static~\cite{tuckerman93} and dynamic~\cite{cao93} properties of 
many-body quantum systems in thermal
equilibrium. It is based on the isomorphism between the
partition function of a quantum mechanical point particle and that
of a classical ring polymer. The number of monomers per ring
is called Trotter number $P$. Recent progress was made concerning
the dynamical interpretation: It was suggested that real
time correlation functions $C(t)$ of observables linear in velocity $v$ 
and/or position $x$ can be calculated exactly with adiabatic PIMD 
methods~\cite{cao96,jang99}. 
The $C(t)$'s follow from the equivalent correlation functions
$C_{\rm c}(t)$ defined for the ring polymer's centroids (center of mass), 
provided that the  $P-1$
inner degrees of freedom are in full thermodynamic equilibrium for
every given position of the centroids.
Furthermore, the characteristic time constant of the centroids' thermostats 
must be large compared to the intrinsic relaxation time of the system.
If these two conditions are satisfied,
$C(t)$'s and $C_{\rm c}(t)$'s Fourier transforms are related through the 
equation 
$\tilde{C}(\omega) = \beta\hbar\omega/2  
( \coth(\beta\hbar\omega/2) + 1) \tilde{C}_{\rm c}(\omega)$, 
where $\beta$ is the inverse thermal
energy. 
In the following, only centroid spectra $\tilde{C}_{\rm c}(\omega)$
will be considered,
as they  reflect directly the density of states.

While we refer to Ref.~\onlinecite{muser02cpc} and to a  future
longer-version publication for more details of our implementation,
we may note that we used a representation of our system that is
diagonal in the harmonic part of the Hamiltonian $\hat{H}$. Moreover,
higher-order corrections to the high-temperature density matrix
were included in our treatment leading to $P^{-4}$ convergence~\cite{kraj02}.
We made sure that the relative systematic errors of the internal energy 
with respect to the classical ground state
due to the use of finite $P$ was less than $10^{-3}$.
While PIMD is a finite-temperature method, it is possible to extract
zero-$T$ behavior by analyzing the convergence of the results
by decreasing $T$, just like the thermodynamic limit can be approximated
from finite-size extrapolations.
We will first consider the most simple
QFK model, which corresponds to a discrete, elastic chain, which is
commensurate with the underlying potential or substrate. 
The Hamiltonian $\hat{H}$ reads
\begin{equation}
\hat{H} = \sum_{n=1}^N {\hat{p}_n^2\over 2m} + {1\over 2} k (x_n-x_{n+1})^2
- V_0 \cos(x_n/b),
\label{eq:hamiltonian}
\end{equation}
where $\hat{p}_n$ and $x_n$ are, respectively  momentum and position 
of particle $n$, $k$ is the  stiffness  of the spring connecting two
neighbored particles, $V_0$ is the coupling strength to the embedding
system, and $2\pi b$ is the substrate's lattice constant.
The periodic boundary condition $x_{n+N} = x_n + 2 \pi b N$
makes the chain commensurate with the substrate.
Our system of units will be defined by  
$V_0$, $b$, 
$\hbar$, and Boltzmann's constant $k_B$. 
Unless otherwise noted, we vary the mass $m$  but leave the
harmonic intrachain coupling $k = 0.1$~$V_0/b^2$ constant.
This value of $k$ is much smaller than the maximun curvature
of the potential $\max(\partial^2_x V(x)) = V_0/b^2$. This challenges
the  continuum approximations, which assume slow
variations of the reduced positions $x_n -2 n\pi b$ with  index $n$.

\begin{figure}
 \includegraphics*[width=8.5cm]{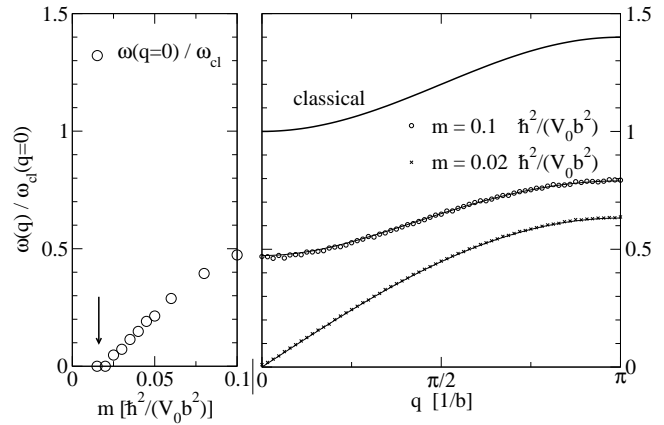}
\caption{
Left: Excitation gap $\omega(q=0)(m)$ at zero wavelength as a function of 
mass $m$.  The arrow indicates where the continuum
sine-Gordon model shows the transition from finite to zero gap.
Right: Phonon dispersion for two different masses and for
the classical case. Lines are fits according to Eq.~(\ref{eq:dispersion})
where $\tilde{k}_0$ is the only fit parameter ($\omega_0 = \sqrt{k_0/m}$).
 \label{fig:dispersion}}
\end{figure}

Classically, the dispersion relation of the chain is simply given by
\begin{equation}
\omega^2(q) = {1\over m} \left[4k\sin^2(qb/2) + \tilde{k}_0\right]
\label{eq:dispersion}
\end{equation}
where $q$ denotes the phonon's wavenumber and $\tilde{k}_0 = k_0 := V_0/b^2$. 
Early calculations~\cite{giachetti85} suggested
that this classical, zero-temperature value is renormalized 
to a reduced effective coupling $\tilde{k}_0$ due to
zero-point and/or thermal fluctuations.
Also for the {\it discrete}, commensurate 1D QFK model, 
Eq.~(\ref{eq:dispersion}) with a non-trivial value for $\tilde{k}_0$
turns out to provide an excellent approximation 
to our calculated phonon spectra. The dispersion relations of our
chain are shown in the right-hand side of 
Fig.~\ref{fig:dispersion} for two different masses.
(The dispersion relation was obtained by fitting Lorentzians
to spectra similar to those shown below in Fig.~\ref{fig:dis_pot}.
In the commensurate system, the lines are much sharper than in the
disordered case.)
The classical zero-temperature dispersion relation is inserted
for comparison.  Each curve requires only one fit parameter,
namely the value for $\tilde{k}_0$. The dispersion relation does
not change when  particle number $N$ and Trotter number $P$ (sufficiently
small but fixed $\beta/P$) are increased. 
We want to note that all relevant imaginary-time correlation 
functions could be well reproduced from the centroid spectra 
$C_c(\omega)$, which made us confident to use
the dynamic interpretation of adiabatic PIMD.

The phonon excitation gap apparently becomes zero at a value 
$m_{\rm c} \approx 0.02$~$\hbar^2/V_0b^2$, as shown in the left-hand side of 
Fig.~\ref{fig:dispersion}.
This in turn implies that sliding can be induced with an arbitrarily small 
external driving force for $m \le m_{\rm c}$. 
As we will show later, the system creeps in the
zero-gap regime when subjected to a small external driving force.
The discreteness of the chain alters the
value of the mass $m_c$ at which the transition from finite gap (no creep) 
to zero gap (creep) takes place. The continuum model predicts this transition
to occur at $m_c =  0.016$~$\hbar^2/V_0b^2$~\cite{colman75}.
Thus for $k = 0.1\,V_0/b^2$, the value for $m_c$ differs 
by about 20\% between the discrete FK and the continuum SG model.
This discrepancy will decrease as the spring stiffness within the chain 
increases as compared to the maximum curvature $k_0 = V_0/b^2$ of 
the embedding potential.
Without showing the data explicitly, we would like to comment again
on the imaginary-time behavior, as this behavior does not rely
on the correctness of the claims made for adiabatic PIMD.
For $m > m_c$, we observe a finite plateau value in the imaginary-time 
correlation function
$G(\Delta n, \Delta \tau) = 
\langle \{x_{n+\Delta n}(\tau+\Delta\tau)-x_{n}(\tau) - 2 \pi b \Delta n\}^2\rangle $
for $\Delta n$ and $\Delta \tau$ $\to \infty$. Similar to what one would
expect from a (quantum) Berezinski-Kosterlitz-Thouless 
transition~\cite{berezinski71,kosterlitz73}, 
$G(\Delta n, \Delta \tau)$ increases logarithmically both with 
$\Delta n$ and $\Delta \tau$ for $m < m_c$.
An accurate determination of $m_c$'s value would yet remain much more 
difficult in terms of an imaginary-time analysis as compared to the one 
based on centroid dynamics.

\begin{figure}
 \includegraphics[width=8.5cm]{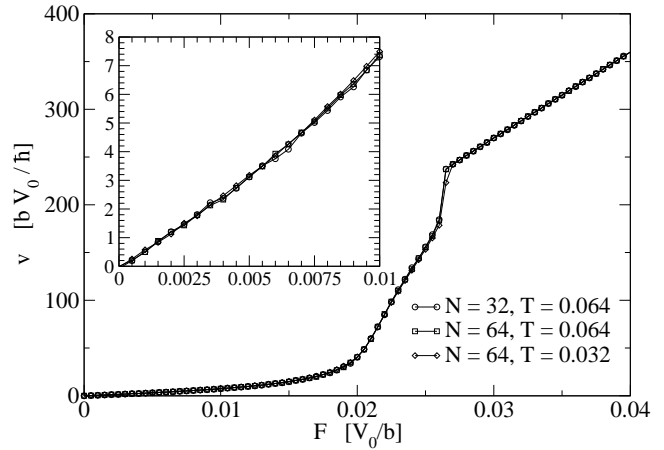}
\caption{Sliding velocity $v$ as a function of externally applied force $F$ 
for mass $m = 0.02$, system sizes $N = 32$ and $N = 64$,
and temperatures $T = 0.064$~$V_0/k_B$ and $T = 0.032$,
starting with a system at rest and slowly increasing $F$.
The inset is an amplification of the small-$v$ regime of the main figure. 
For $F < 0.01$, increasing or decreasing the
thermostat (i.e., the externally imposed damping)
by a factor of two has almost no effect on the $v(F)$ curve.
 \label{fig:force_vel}}
\end{figure}

To study the response to an externally applied field, a homogeneous
force $F$ is applied to each particle by adding a term $-F \sum_n x_n$ to
$\hat{H}$ in Eq.~(\ref{eq:hamiltonian}). In Fig.~\ref{fig:force_vel}
the FK chain's sliding velocity is shown as a function of the driving force
for $m = 0.02$. This is the value of $m$, where the gap closes.
It is found that the response in $v$ is linear with $F$ at very small
and very large $F$ with different values for the effective damping
coefficient $\gamma_{\rm eff} = F/v$. 
While (quantum) continuum approximations predict $\gamma_{\rm eff}$ to be 
zero~\cite{zotos03}, the chain's discreteness is known 
to change this property in classical systems, because kink-phonon
interactions damp solitons~\cite{combs83}.
Interestingly, $\gamma_{\rm eff}$
is independent of temperature $T$ as $T$ approaches zero in both linear 
regimes. This rules out the possibility that thermal fluctuations assist 
the system to overcome energy barriers at small $F$.  
Our observation is yet in partial contrast to the behavior of Luttinger 
liquids, which show 
$v/F \propto T^{\alpha}$ with $\alpha > 0$~\cite{ando98}. 

Another similarity with the dynamics of the classical (discrete)
FK model is seen at an external force $F = 0.02$, where the 
friction-velocity relation exhibits a cusp, separating 
a low-friction, low-velocity from 
a high-friction, high-velocity regime~\cite{combs83}.
Finally, at external forces $F > 0.025$, the response
crosses over into a completely unpinned regime, in which the motion
of the FK chain is mainly opposed by the drag coefficient associated
with the heat bath. In that regime, soliton-related dissipation
mechanisms become unimportant.
Transport of chains consisting of beads with masses $m > m_c$ requires
either finite temperatures and/or finite forces. For example
for $m = 5 m_c$, our numerical data provides an upper bound for the
mobility or inverse effective damping 
$\gamma_{\rm eff}^{-1}(m=0.1) < 2\cdot 10^{-6} \gamma_{\rm eff}^{-1}(m=0.02)$.

The previous analysis shows that PIMD can distinguish well between
zero gap and finite gap. It is now possible to generalize the
treatment for which the quantum mechanical ground state (of the
continuum model) is not known analytically. 
One of the topics of current scientific interest 
is the interplay of disorder and quantum 
fluctuations~\cite{gorokhov02}. 
Here we want to investigate the effect
of disorder on the spectral properties of the quantum FK chain.
This is done by replacing the external potential $V_0\cos(x_n/b)$
in Eq.~(\ref{eq:hamiltonian}) with a random potential $U_H(x)$,
which is constructed in the following way: We add patches of the
functional form $V_0\cos(x/b)$ of length $\pi b$ where
the underlying domain is chosen randomly to be either $[0 ,\pi b]$
or $[\pi b , 2 \pi b ]$. The patches are shifted by a constant
so that no discontinuity in the potential occurs.
This leads
to $\langle \{ V(x+\Delta x) - V(x) \}^2 \rangle \propto \Delta x ^{2H} $
on scales $b << \Delta x << \pi b N$.
The corresponding surface roughness exponent $H$ takes the value $H = 1/2$.
In order to construct a random potential with $H=0$,
the potential is either zero on a length of $ \pi b$ or - 
with same probability - it takes the functional form of
$V_0 \{1+\cos (x/b)\}$
on the interval $-\pi \le x/b < \pi $. 
Our numerical analysis of the results for the case $H=0$  suggests a 
finite gap for a mass $m > m_c^{(H=0)}$ and zero gap for $m < m_c^{(H=0)}$,
in agreement with the predictions by Gorokhov et al.~\cite{gorokhov02}. 
For the model potential under 
consideration, one phonon branch is observed at very large $m$,
which then forms a broad band for masses larger than but in the order of $m_c$.
At $m < m_c$, there is only one relatively narrow branch, which can 
be well described with Eq.~(\ref{eq:dispersion}) and $\tilde{k}_0 = 0$.

The situation is more complex for disordered potentials with roughness
$H = 1/2$.
In particular, we find that classical and quantum system behave
qualitatively different. In agreement with the predictions by 
Chauve et al.~\cite{chauve98epl}, there is no indication for
the classical mobility to become zero at finite temperatures. 
In contrast, the quantum mechanical mobility always appears to
be zero for $H=1/2$, no matter how small $m$ is. Whenever, we were prompted
to believe the quantum $H = 1/2$ system was depinned, doubling the system size 
dramatically reduced the velocity-velocity spectrum $C(q,\omega)$
near zero wavelengths and zero frequencies. 
The zero-wavevector spectrum $C(q=0,\omega)$
for classical and quantum mechanical FK chains is shown in the left-hand 
side of Fig.~\ref{fig:dis_pot}, where the qualitative difference of
$C(q,\omega)$ in the limit $(q,\omega) \to (0,0)$ is bourne out.
(Note that due to statistical uncertainties and finite system size,
one can never observe $C(0,0) = 0$ in a computer simulation.)

\begin{figure}
 \includegraphics[width=8.5cm]{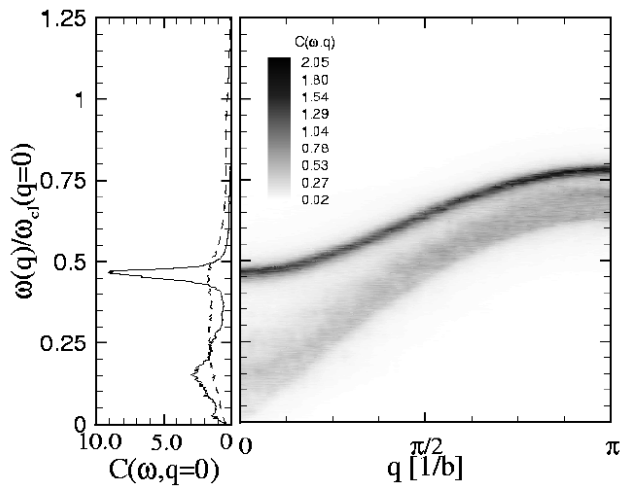}
\caption{Left: Zero wavevector spectrum for the quantum ($m = 0.02$, thick line)
and classical ($k_B T = 1.635$~$V_0$, thin dashed line)  FK model embedded
in a disordered potential of roughness $H = 1/2$. $T$ was chosen such
that the average kinetic energy was the same in the classical as in the
quantum case. Right: Full spectrum of the velocity correlation
function for the quantum system. 
Average over 12 different disorder realizations. 
}
 \label{fig:dis_pot}
\end{figure}

The spectrum at finite wavelengths of the quantum system, which is
shown on the right-hand-side of Fig.~\ref{fig:dis_pot} shows two
phonon branches. The figure is the average over 12 different disorder
realizations. A single realization shows much sharper lines. It
appears that the system is not self-averageing.

In conclusion, we find that the regular zero-temperature, {\it discrete},
quantum sine Gordon model has an insulating  region with $m > m_c$ and
finite gap, while for $m \le m_c$ the gap closes resulting in finite
resistance.  Similar behaviour is found for a quantum chain on a disordered
potential with roughness exponent $H = 0$. However, $H = 1/2$ automatically
leads to a zero-temperature insulator, even though small thermal 
fluctuations are sufficient to induce creep.\nopagebreak

\newpage

\begin{acknowledgments}
Support from NSERC, SHARCNET, the BMBF through Grant 03N6015, and
from the Materialwissenschaftliche Forschungszentrum  
Rheinland-Pfalz is gratefully acknowledged.
We thank Kurt Binder and Stefan Kehrein for useful discussions.
\end{acknowledgments}

\end{document}